\documentclass[letterpaper,superscriptaddress,aps,twocolumn]{revtex4-2}
\usepackage{amsmath,amssymb}
\usepackage{bm}
\usepackage{graphicx,color,natbib}
\usepackage{placeins}
\usepackage[dvipsnames]{xcolor}
\usepackage{siunitx}
\usepackage{oplotsymbl}

\begin{document}

\title{The Free-Energy Barrier of Precritical Nuclei in Hard Spheres is Consistent with Predictions}

\author{Lars K\"{u}rten}
\affiliation{Gulliver UMR 7083, CNRS, ESPCI Paris, Universit\'{e} PSL, 75005 Paris, France}
\author{Antoine Castagn\`{e}de}
\author{Frank Smallenburg}
\affiliation{Laboratoire de Physique des Solides, CNRS, Universit\'{e} Paris-Saclay, 91405 Orsay, France}
\author{C. Patrick Royall}
\affiliation{Gulliver UMR 7083, CNRS, ESPCI Paris, Universit\'{e} PSL, 75005 Paris, France}

\begin{abstract} 
Predicting crystal nucleation  is among the most significant long--standing challenges in  condensed matter. In the system most studied (hard sphere colloids), the comparison between experiments performed using static light scattering and computer simulations is woeful, with a discrepancy of over 10 orders of magnitude. The situation with other well-studied materials (such as water and sodium chloride) is no better. It has thus far proven impossible to access the regime of this discrepancy with particle-resolved techniques which might shed light on its origins, due to the relatively sluggish dynamics of the larger colloids required for confocal microscopy.

Here we address this challenge with two developments. Our work is a marked improvement in the precision of mapping the state point of experiments to simulation. For this, we employed a combination of novel machine-learning methods for particle tracking and higher-order correlation functions. Our second innovation is to consider the free energy of \emph{pre--critical} nuclei which can be detected in the discrepancy regime. These are in agreement with computer simulation. This is the first time that such free energies  have been successfully compared between experiment and simulation in any material as far as we are aware. The agreement provides important validation of rare event sampling techniques which are used very widely in simulation, but which can seldom be directly compared with experiment. 
\end{abstract}

\maketitle

\section{Introduction}
Crystal nucleation is a truly everyday phenomenon with very wide--ranging industrial and environmental implications from purification of chemicals to crystallisation of aerosol droplets being a major uncertainty in climate change modelling~\cite{sosso2016,ipcc2023}. It is clearly important to be able to make meaningful predictions of the rate of  nucleation of crystallites. Despite very recent advances in atomistic materials in a limited range of nucleation rates~\cite{moller2024}, the comparison of nucleation rates between prediction and experiment is spectacularly bad, in well-studied materials such as water~\cite{sosso2016}, and in the case of NaCl, there is even significant discrepancy between different implementations of the same computational model~\cite{finney2023}.

In such general challenges, model experimental systems, such as colloids, play an important role, as their classical interactions make accurate large-scale computer simulations possible, while their kinetics are convenient from an experimental perspective~\cite{evans2019,royall2024}. The kinetics are controlled by the Brownian time taken by an isolated sphere to diffuse by its own radius $\tau_B = 3\pi\eta\sigma^3/8 k_B T$ where $\eta$ is the solvent viscosity.

\begin{figure}[!hb]
\centering
\includegraphics[width=.8\linewidth]{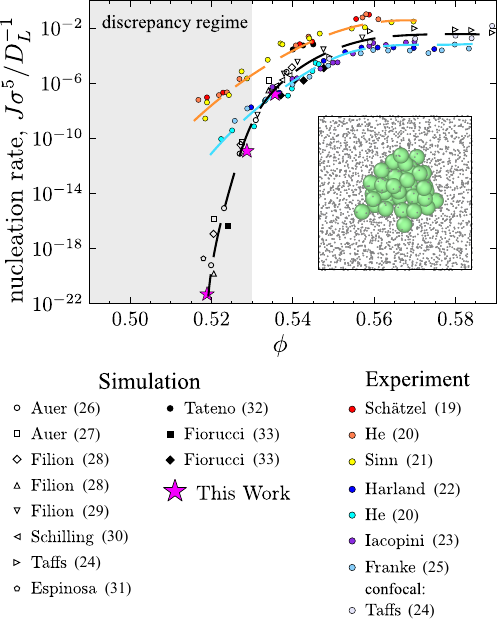}
\caption{\textbf{Nucleation rate discrepancy and pre-critical nuclei.}
Reduced nucleation rates from experiments \cite{schatzel1993,he1996,sinn2001,harland1995,iacopini2009,taffs2013,franke2014} and computer simulations \cite{auer2001prediction,auer2001suppression,filion2010,filion2011jcp,schilling2011jpcm,taffs2013,espinosa2016,tateno2019,fiorucci2020}. Experiments are divided into two branches. Upper branch (orange) shows results of experiments without density matching, lower branch (blue) with approximate density matching and almost no sedimentation. Inset shows rendering of coordinates from experimental system ($\phi=0.54$) with pre-critical nuclei (green) identified with $q_6$ bond-orientational order parameters of Ten Wolde~\cite{wolde1996simulation}. Liquid particles (grey) are depicted smaller.}
\label{figNucleationRates}
\end{figure}

However, in the most studied system, colloidal hard spheres, as Fig.~\ref{figNucleationRates} shows, the discrepancy between experiment and simulation is wild ~\cite{royall2024,auer2004,palberg2014,filion2010crystal}.  This discrepancy is important beyond the immediate application of nucleation because the rare event sampling computational methods required here are used across the molecular sciences from physics to biochemistry ~\cite{allen2009forward,kastner2011umbrella, bolhuis2021transition}. Direct comparison of rare event sampling is seldom possible with the level of detail available here, which renders solution of this large discrepancy all the more urgent.

In colloidal hard spheres, the nucleation rate vanishes at the freezing boundary and increases dramatically upon raising the chemical potential (or volume fraction). Here, two regimes can be identified as shown in Fig.~\ref{figNucleationRates}. At high supersaturation, where nucleation is rapid, experiment and simulation agree reasonably well. At low supersaturation where nucleation is very slow, a huge discrepancy emerges as shown by the shading in Fig.~\ref{figNucleationRates}. The experiments in this \emph{discrepancy regime} were carried out using light scattering in which the number and size of the nuclei is inferred indirectly~\cite{palberg2014,royall2024}. More recent work using confocal microscopy where the coordinates of the particles can be tracked, offers the chance to measure more precisely the nuclei that form~\cite{gasser2001} and to compare directly with predictions from simulation. Alas, while such data offers impressive insight~\cite{tan2014}, it comes at a high price: colloids that are large enough to track (typically 2-3 $\mu$m in diameter) have a Brownian time $\tau_B\sim 10$s, rendering rare events like nucleation in weakly supersaturated suspensions inaccessible on reasonable experimental timescales.

There are two possible reasons for the discrepancy: either the nucleation rate is different between the experiments and that predicted by computer simulation, or the state point is not mapped correctly between the two. Until recently~\cite{kale2023}, accurate determination of the state point, i.e. the volume fraction $\phi$ of an experimental colloidal system has been very difficult~\cite{poon2012,royall2013myth}. We note that a mechanism has been put forward which suggests that the light scattering measurements may overstate the number of nuclei~\cite{wohler2022}, although whether this could explain the vast discrepancy remains unclear~\cite{royall2024}.

Here we address the nucleation discrepancy with two developments using real-space studies of colloids using confocal microscopy. Firstly, we consider the nucleation barrier. In classical nucleation theory (CNT), the nucleation rate 
\begin{equation}
J = \kappa e^{-\beta \Delta G^\ast}
\label{eqJ}
\end{equation}
where the prefactor $\kappa$ includes both the attempt frequency of attaching/detaching a  particle from the critical nucleus  (or jump frequency) per unit volume, and the Zeldovich factor which accounts for the probability that a critical nucleus will grow out to become a macroscopic nucleus~\cite{kalikmanov2012classical}. The attempt frequency is reasonably well-matched between experiment and prediction. In any case, it does not vary enough numerically to account for the magnitude of the discrepancy. We therefore focus on the barrier height. Although confocal microscopy can determine critical nuclei in the regime where nucleation is rapid~\cite{taffs2013, ketzetzi2018}, here our focus is on the discrepancy regime where critical nuclei are large and exceedingly rare. Therefore, we seek to measure the free energy of formation of smaller, pre-critical nuclei, whose occurrence is sufficiently frequent to be accessible in experiments using confocal microscopy.

Secondly, we aim to map state points between experiment and simulation with hitherto unprecedented precision. To do so, we improve the particle tracking with recently developed methods which implement machine learning~\cite{kawafi2024} and use higher-order correlation functions which are more sensitive to volume fraction than pair correlation functions such as the radial distribution function $g(r)$. We calculate the effective free energy as a function of (pre-critical) nucleus size from their concentration in supersaturated fluids. We find good agreement with computer simulations when taking into account the remaining experimental errors. This is the first time that the free energy of formation of nuclei has been successfully compared between experiment and simulation in any material as far as we are aware.  Within the accuracy of our experiments, we find no discrepancy with respect to our hard sphere simulations.

\section{Brief Methodology}
\label{sectionMethods}
\textit{Experiments. --- } 
We used sterically stabilized PMMA spheres with a diameter of 2.0 microns and polydispersity of 4\%~\cite{wu2025}. The particles are dispersed in refractive index and density matching solvent mixtures with salt added to screen electrostatic interactions.

To extract particle coordinates from the confocal images we used a deep learning approach called `‘Colloidoscope'’~\cite{kawafi2024}. This particle tracking routine was explicitly developed to track colloidal particles in crowded environments where conventional approaches based on the Crocker-Grier algorithm~\cite{crocker1996methods}, for example, reach their limits. ``Colloidoscope'’ allowed us to access the coordinates of around 98\% of the particles in the sample with a tracking error of 5\% of the particle diameter.

\begin{figure*}[!ht]
\centering
\includegraphics[width=0.9\textwidth]{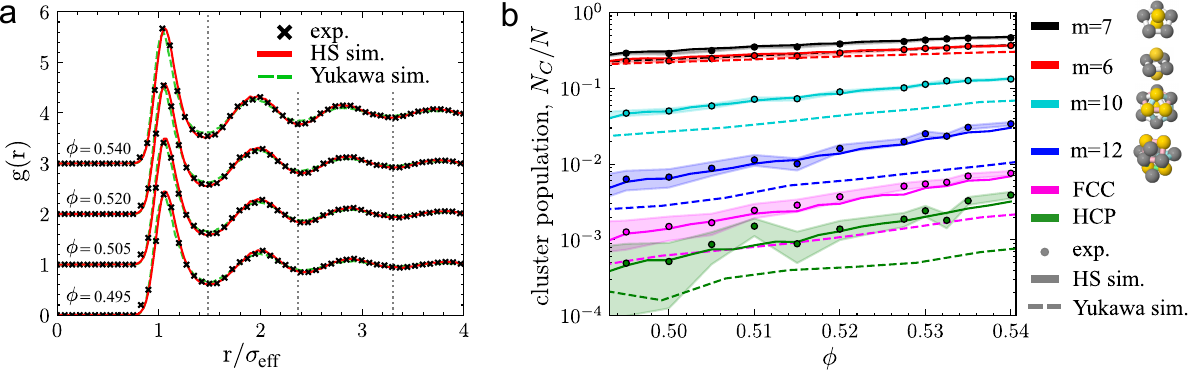}
\caption{\textbf{Mapping between experiment and computer simulation at fluid state points around the coexistence regime.}
(a) Radial distribution functions at a range of volume fractions. Data are offset for clarity.
(b) Higher-order correlations in the hard sphere fluid as a precise probe of state point matching. Selected clusters are plotted as a function of volume fraction. Errors were calculated from the standard deviation of mean populations of different experiments mapped to the same volume fraction.}
\label{figMapping}
\end{figure*}

\textit{Computer Simulations. --- }
In order to compare our experimental results to ideal hard spheres, we perform both unbiased and biased simulations of polydisperse hard spheres. We match the experimental system with a Gaussian distribution with $4\%$ polydispersity. For the unbiased simulations, we use event-driven molecular dynamics simulations~\cite{smallenburg2022} of $N = 10^5$ particles in which we monitor the distribution of sizes of crystalline nuclei 
using the $q_6$ bond-orientational order parameters~\cite{wolde1996simulation}.

We also perform biased Monte-Carlo simulations in the isobaric-isothermal ensemble ~\cite{frenkel2023understanding}, using the number of particles $n$ in the largest crystalline nucleus as our bias parameter, and relying on the same method for nucleus identification as in the unbiased simulations. The umbrella sampling simulations allow us to reconstruct the free-energy barriers as a function of nucleus size at different supersaturations from which we can obtain the nucleation rate. As an additional point of comparison, we also performed simulations of particles interacting via a hard-core Yukawa interaction, in order to test for the effects of residual electrostatic interactions between the particles.

\textit{Mapping Experiments to Hard Sphere State Points. --- }
To map the state point between the experiment and simulation, we proceed as follows. As the colloids have some degree of softness~\cite{royall2013myth}, we seek the \emph{effective} hard sphere volume fraction $\phi_\mathrm{eff}$. We compare the total pair correlation function $h(r)=g(r)-1$. This is shown in Fig.~\ref{figTotalCorr}. This provides a first estimate of the true effective volume fraction $\phi^\mathrm{true}_\mathrm{eff}$. When we count the particles, since some are not tracked, we obtain a smaller volume fraction $\phi^\mathrm{count}_\mathrm{eff}  \leq \phi^\mathrm{true}_\mathrm{eff}$.

Our second comparison is to use the population of certain clusters with a low free energy~\cite{robinson2019} identified with the topological cluster classification (TCC)~\cite{malins2013tcc} as shown in Fig.~\ref{figMapping}(b). We weight the TCC populations and the $h(r)$ comparisons to arrive at an optimal mapping. Further details are provided in Appendix A.

\section{Results}
\label{sectionResults}
We compare the experimental and simulated state points using the radial distribution function $g(r)$ in Fig.~\ref{figMapping}(a). The agreement is very good indicating that the state point is well matched. This is replicated in the total correlation function shown in the Appendix.

However, we should note that  $g(r)$ is not a strong function of volume fraction in this regime. It is desirable to have a more precise probe of volume fraction and here we turn to higher-order correlations. It is possible to determine populations of clusters which minimise the local free energy of hard spheres~\cite{robinson2019} and the population of these is shown in Fig.~\ref{figMapping}(b). Overall, the agreement between experiment and hard-sphere simulation is very good, providing further evidence that the state point has been well matched. The hard-core Yukawa simulations have a a noticeably lower population of the larger cluster sizes ($m \geq 10$) and hence are not compatible with our experimental observations.

\textit{Free energy of precritical nuclei. --- }
Having mapped the state point between experiment and simulation, we now move on to consider the main result, the effective free energy of formation of pre-critical nuclei, $\Delta G(n)/k_BT$ where $n$ is the observed number of particles in the nucleus. Here we determine the free energy of formation 
\begin{equation}
\Delta G(n)/k_BT=-\log(-N_n/N)
\label{eqDGn}
\end{equation}
where $N_n$ is the number of observed nuclei of size $n$ and $N$ is the total number of particles sampled. As noted above, here we use the $q_6$ bond-orientational order parameters to determine the nucleus size, and -- for the simulations -- include the tracking error and random particle deletion in our analysis. In other words, our reaction coordinate $n$ here is the \textit{observed} nucleus  size $\Delta G_\mathrm{obs}$, including the effects of imperfect particle tracking, rather than the \textit{bare} nucleus size that one would obtain from applying the bond-orientational order parameters to the true particle coordinates (which we can only do in simulations).

\begin{figure*}[!t]
\centering
\includegraphics[width=0.8\textwidth]{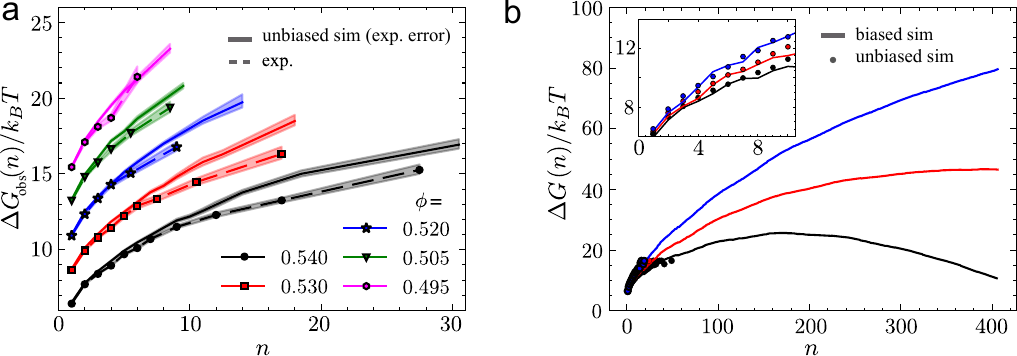}
\caption{\textbf{Free energy of precritical nuclei.}
(a) Comparison of the start of the observed nucleation barrier calculated from size distribution of precritical nuclei from experiments and unbiased hard sphere computer simulations with experimental error ($\Delta G_\mathrm{obs}$). Data points for both were binned into different nuclei sizes to compensate for fluctuations in the number of identified nuclei. 
The error bars are calculated as the square root of the number of nuclei 
in each bin. Data are offset by $2k_{B}T$ each for clarity.
(b) Comparison of nucleation barriers from biased hard sphere computer simulations (solid line) and nucleation barriers calculated from precritical nuclei in unbiased hard sphere computer simulations (data points), both without experimental error ($\Delta G$). Inset shows zoomed in start of the barriers.}
\label{figBarriers}
\end{figure*}

In Fig.~\ref{figBarriers} we show the pre-critical behavior of the observed nucleation barrier for four representative state points in the coexistence regime. Further state points are shown in the SI. For the lowest volume fractions, the agreement between experiment and simulation is generally very good. For $\phi=0.495, 0.505, 0.520$, we can identify no meaningful disagreement between experiment and simulation. As Fig.~\ref{figNucleationRates} shows, this accounts for much of the regime of the nucleation rate discrepancy.

For the highest two volume fractions $\phi=0.530$ and $0.540$, there is a small difference between experiment and simulation of around $k_BT$. We emphasise that this is very much less than that implied in Fig.~\ref{figNucleationRates}. Indeed, all other contributions being equal, the discrepancy in nucleation rates implied by a difference of $k_BT$ is merely a factor of $e \simeq 2.718$, which is so small as to be irrelevant in the context of the nucleation discrepancy (Fig.~\ref{figNucleationRates}). And in any case, this emerges in a higher volume fraction regime $\phi \geq 0.5300$ rather than that of the nucleation rate discrepancy.

\begin{figure*}[!t]
\centering
\includegraphics[width=0.8\textwidth]{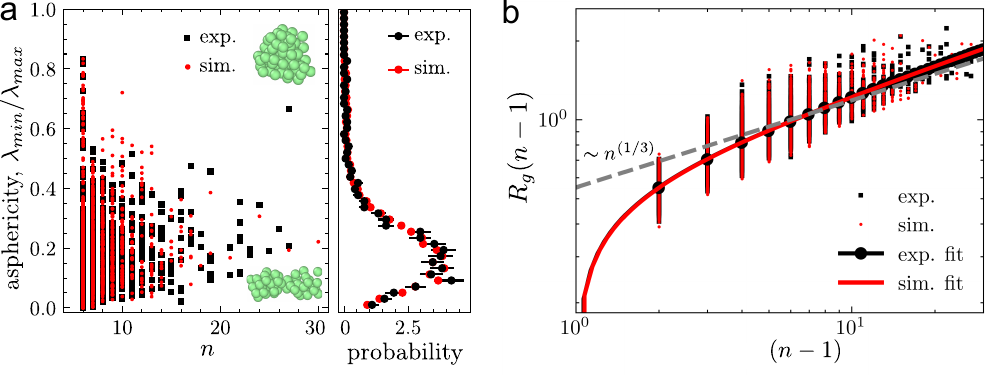}
\caption{\textbf{Asphericity and radius of gyration}
(a) Asphericity quantified by ratio of minimum and maximum variance along different axis of the pre-critical nuclei for hard sphere simulations and experiments.
(b) Radius of gyration as a function of the nucleus 
size. Fit of shape $R_{g}(n-1)=a(n-1)^{b}$ for experiments: $a=0.552\pm0.005, b=0.361\pm0.006$ and computer simulations: $a=0.550\pm0.005, b=0.363\pm0.006$. }
\label{figPCA}
\end{figure*}

\textit{Properties of precritical nuclei. --- } 
Our study naturally provides a large set of crystal nuclei particle coordinates, along with those of the surrounding fluid. We now proceed to analyze the properties of these nuclei. In Fig.~\ref{figPCA}(a) we show the asphericity of the nuclei. The calculation is based on a simple Principal Component Analysis and is discussed in the Materials and Methods section. In terms of the asphericity, experiment and simulation are very well matched. Larger nuclei tend to be more spherical.

Another important geometric quantity is the compactness of the nuclei. This has been related to the crystal-fluid interfacial free energy~\cite{taffs2016}. Like previous work~\cite{taffs2016} we see a considerable spread in the values of the radius of gyration for a given nucleus size. We emphasize the agreement between experiment and simulation here. We also analyzed the influence of a change in volume fraction on the averaged radius of gyration. Fig.~\ref{figPowerLaw} shows the parameters of the fitted power law as a function of the state point. While the data show considerable fluctuations, and more statistics would be needed before strong conclusions can be drawn, there is some evidence for a drop in the fitted power law $b$ with volume fraction (see Fig.~\ref{figPowerLaw}). This could be interpreted as nuclei becoming less compact with increasing volume fraction, though we caution that this observation should be treated as preliminary.

\section{Discussion}
We have measured the free energy of small crystalline nuclei in colloidal hard spheres. Within our measurement error, we see no significant difference between experiment and simulation. Of course, our data do \emph{not} measure the height of the critical nucleus barrier $\Delta G^*$  and thus some caution should be exercised in our results. It is possible that the small discrepancy that we find between experiment and simulation could increase for larger nuclei. Nevertheless, is it still tempting to speculate as to why in our results the discrepancy has all but disappeared.

There are two key differences with respect to the earlier experiments. Firstly, we have used real-space confocal microscopy to measure the nuclei. This means that the analysis of coordinate data is essentially identical between experiment and simulation, with few assumptions being made. Secondly, we have matched the state point to the simulations carefully, by exploiting our real space data.

While the methodology to detect nuclei is of course very different with respect to the earlier experiments, we believe that the methods used to detect nuclei are broadly sound (see Ref.~\cite{schope2007} for a description of the scattering methods to determine nuclei size). This brings us to consider the second factor, mapping the state point between experiment and simulation. While hard spheres are of course a model system, and are well-controlled, accurate determination of the volume fraction is fraught with difficulty~\cite{poon2012,royall2013myth}. We believe that the method we have implemented here marks an improvement over previous determinations of the volume fraction in nucleation studies which used a determination of the phase diagram to infer the (effective) volume fraction. The method developed by Paulin and Ackerson~\cite{paulin1990} which underpinned the determination of $\phi$ in the earlier studies~\cite{royall2024} is undoubtedly ingenious, but it does suffer from competing timescales of nucleation, phase separation and sedimentation, which may couple to influence the value of $\phi$ obtained.

While we emphasize that the agreement we obtain in mapping of volume fraction between experiment and simulation is very good, some discussion of the error is clearly in order. We weight the mapping between that of the total correlation function $h(r)$ (Fig.~\ref{figTotalCorr}) and the higher order correlations expressed through the cluster population. Now if we take the absolute best agreement for $h(r)$ (expressed through the least squares difference between the experiment and simulation plots) then in fact we end up with a value $\delta\phi=0.0025$ less than that we use. This lower volume fraction has a considerably worse agreement with the TCC clusters. However, when we plot $h(r)$ for the slightly higher volume fraction, there is no discernible change. We therefore take our error in volume fraction to be $<0.0025$, which is an order or magnitude better than that estimated previously~\cite{poon2012} and indeed close to the limit that can reasonably be expected with experiments on model colloids~\cite{royall2024}.

\section{Materials and Methods}
\textit{Experiments. --- } 
\label{sectionMethodsExperiments}
We investigated sterically stabilized PMMA spheres with a diameter of 2.0 microns and polydispersity of 4\% as determined using static light scattering. The particles are dispersed in two different solvent mixtures. One consisted of cyclohexyl bromide (CHB) and cis-decalin and the other one of CHB, tetralin and decalin. In both 4 mMol of tetrabutyl ammonium bromide (TBAB) salt was dissolved to screen residual charges on the surface of the particles and avoid softness in the interaction potential. By comparing the radial distribution function to hard sphere computer simulations, we found that our particles interact very similar to the hard sphere interaction and that the choice of the solvent components has no measurable influence on the outcome of our analysis.

Nevertheless, we favour the 3-component mixture because it enables us to match the density and refractive index of the particles separately. We exclusively investigated samples with carefuldensity matching to avoid any influence of sedimentation on the nucleation events~\cite{russo2013,wood2018}. The PMMA spheres were left in the solvent for a week to equilibrate, loaded to a glass capillary and then imaged by 3D confocal laser scanning microscopy to obtain particle-resolved information.

\textit{Particle Tracking. --- } 
\label{sectionMethodsParticleTracking}
The investigation of size distributions and structure of pre-critical nuclei places great demands on the quality of our experimental data. Missed particles and positions with a large localisation error not only lead to a higher uncertainty in the effective volume fraction but could also shift the calculated nucleation barrier. Therefore to extract particle coordinates from the experimental image we use a recently developed deep-learning routine called ``Colloidoscope'’~\cite{kawafi2024}. Especially in the regime of high volume fractions where particles are densely packed and their intensity distributions overlap, ``Colloidoscope'’ tracks a higher number of particles with a smaller localisation uncertainty in comparison to conventional approaches, based for example on the Crocker-Grier-Algorithm~\cite{crocker1996}. We compared the outcome of different tracking routines by analysing the number of tracked particles, the shape of the $g(r)$ and the population of higher-order clusters. We found that ``Colloidoscope'’ tracks more particles, produces a higher first peak in the $g(r)$ and an increased higher-order cluster population, which indicates that the particles are tracked with better precision. In addition ``Colloidoscope'’ is particularly robust against the influence of photo bleaching, allowing us longer exposure times and higher resolution for a longer period of time. From comparison to computer simulations we found that ``Colloidoscope'’ tracks approximately 98\% of the particles in the experimental image with an error in position of around 5\% of the particle diameter.

\textit{Computer Simulations. --- } 
\label{sectionSimulations}
In order to compare our experimental results to ideal hard spheres, we perform both unbiased and biased simulations of size polydisperse hard sphere mixtures. We match the experimentally observed distribution of particle diameters by employing a deterministically generated Gaussian distribution with $4\%$ polydispersity (defined as the ratio of the standard deviation of the particle size $\sigma$ over the mean particle size $\bar{\sigma}$). For the unbiased simulations, we use event-driven molecular dynamics simulations~\cite{smallenburg2022} of $N = 10^5$ particles starting from a fluid phase, for a range of different volume fractions $\phi$ in the supersaturated regime.

To facilitate comparison to the experimental data, we also recalculate the expected distributions of 
sizes of nuclei and TCC cluster concentrations in the presence of experimental errors. In order to mimic the experiments as closely as possible, we include the effects of a finite imaging volume, error in the position of the particles, and ``missed'' particles during tracking. First, we divide each simulation box of $N = 10^5$ particles into sub-blocks of equal volume containing approximately $N_\mathrm{sub}= 5000$ particles in order to simulate the effects of a finite imaging volume on the detection of nuclei. Second, we apply a random displacement to all particle positions, drawn from a Gaussian distribution with zero mean and a standard deviation of $d_\mathrm{err} = 0.05$. Third, we delete a small fraction $\alpha$ of the particles from the configuration (typically less than $2\%$), with the particles chosen at random. Note that this approach does not take into account the possibility of correlations between tracking error and particle environments, which could in principle be present in the experimental data.

We also perform biased Monte Carlo simulations in the isobaric-isothermal ensemble with an umbrella sampling scheme~\cite{frenkel2023understanding}, using the number of particles $n$ in the largest crystalline nucleus as our bias parameter. The nucleus size was determined based on standard bond-orientational order parameters, described in the following subsection. Note that the role of the bond-orientational order parameters on the height and shape of the free-energy curve was explored in Ref.~\cite{filion2010crystal}. The umbrella sampling simulations allow us to reconstruct the free-energy barriers as a function of nucleus size at different supersaturations.

We also considered a hard-core Yukawa potential, taking into account softness from the screened residual charges on the surface of the particles. The pair potential is given by
\begin{align}
\beta u(r)=
\begin{cases}
\beta\epsilon\frac{\exp[-\kappa\sigma(r/\sigma-1)]}{r/\sigma}, &r>\sigma\\
\infty, & r<\sigma
\end{cases},
\end{align} 
where $\beta\epsilon$ is the value of the contact potential, $\kappa$ denotes the inverse Debye screening length and $\sigma$ is the particle hard-core diameter. In this work we chose $\beta\epsilon=11$ and $1/\kappa\sigma=0.05$ and mapped the state point of the system with softer interaction potential on hard sphere systems with the Barker-Henderson equation~\cite{barker1976}.

\textit{Order Parameter. --- } 
\label{sectionMethodsOrderParameter}
We monitor the distribution of sizes of crystalline nuclei using the $q_6$ bond-orientational order parameters of Ten Wolde \textit{et al.}~\cite{wolde1996simulation}, for which we employ the dot-product cutoff value to consider pairs of particles to be connected to be  $q_6(i) \cdot q_6(j) = 0.7$, and where we consider a particle to be in a crystalline environment when its number of solid-like neighbors is at least 5. Specifically, we determine nearest neighbors using the SANN algorithm~\cite{van2012parameter} which avoids the need for another cutoff parameter.

\textit{Detecting Higher-Order Structures. --- } 
\label{sectionMethodsTCC}
We analyse our data with the topological cluster classification (TCC)~\cite{malins2013tcc}. This detects geometric motifs whose bond topology is identical to that of clusters which minimise the local free energy for hard spheres in dense fluids~\cite{robinson2019}.

The TCC uses Voronoi tessellation to identify nearest neighbours and then detects shortest path rings (3,4,5 particles) in those configurations. Basic clusters are constructed on them and by adding more particles clusters of various size can be identified. We use identical methods for analysing experiments and computer simulations to prevent biasing certain cluster type populations.

\textit{Asphericity Calculation. --- } 
\label{sectionMethodsSphericity}
We use Principal Component Analysis (PCA) to quantify the sphericity of crystal nuclei. We apply PCA on sets of 3D coordinates in a crystal arrangement to find the two axes of our crystal nuclei with the highest ($\lambda_\mathrm{max}$) and lowest variance ($\lambda_\mathrm{min}$). Here $\lambda$ denotes the eigenvalues of the covariance matrix. For a perfect spherical nucleus, the two variances should have a similar value and therefore $\lambda_\mathrm{min}/\lambda_\mathrm{max}$ should be close to one. For our data we find distributions with a peak at 0.2 in computer simulations and experiments. This is likely because we only consider relatively small numbers of particles per pre-critical crystal nuclei. We also analysed bigger crystal nuclei of up to 100 particles that were produced with biased computer simulations and found that the ratio of variances slowly converges towards one with increasing nucleus size.

\begin{acknowledgments}
It has been a pleasure to discuss the challenge of the hard sphere nucleation problem with
Marjolein Dijkstra,
Daan Frenkel,
Peter Harrowell,
Laura Filion,
John Russo
and
Thomas Speck.
The authors gratefully acknowledge the Agence National de Recherche grant DiViNew for support. This work benefited from the technical contribution of the joint service unit CNRS UAR 3750.
\end{acknowledgments}

\newpage
%

\newpage
\appendix
\onecolumngrid  
\newpage
\section{Mapping the State Point Between Experiment and Simulation}
Nucleation barriers are highly sensitive to the state point of the system. We want to calculate the start of the barrier from the size distribution of precritical nuclei found in experiments and compare to computer simulations. Therefore, special attention must be paid to the determination of effective volume fractions in experiments. In the following a newly developed procedure of carefully mapping experimental data sets onto computer simulations is described. One advantage of this technique is that almost no \emph{a priori} knowledge of the properties of the sample is required for the complete characterisation. An experimental data set of a hard spheres system is described by a few parameters. These include the polydispersity, the effective volume fraction, the localization uncertainty, and the fraction of particles that get lost during the tracking process.

The polydispersity was determined from static light scattering experiments. The localisation uncertainty of the tracking routine can be inferred from its effect on the shape of the first peak of the radial distribution function and was determined in a previous publication~\cite{kawafi2024}. We found that adding a Gaussian error with 5\% standard deviation to the simulation coordinates matches the experimental conditions.

\begin{figure}[!b]
\centering
\includegraphics[width=.55\textwidth]{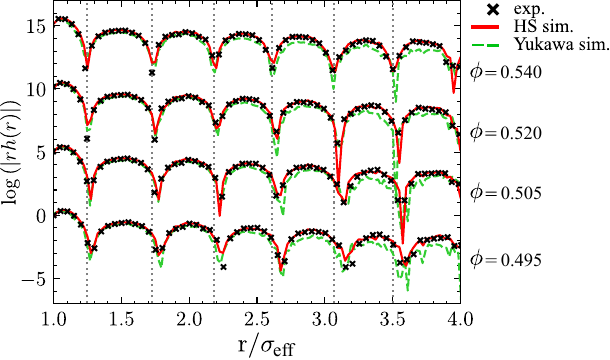}
\caption{\textbf{Total Correlation Function}. Total correlation function at a range of volume fractions in coexistence regime. Data are offset for clarity.}
\label{figTotalCorr}
\end{figure}

For the mapping we use two observables, both sensitive to the state point of the system and the interaction potential of the particles. The first one is the logarithm of the total correlation function $\log(|rh(r)|)$ with $h(r)=g(r)-1$ (Fig.~\ref{figTotalCorr}). The decay of the height of the peaks is determined by the state point of the system. The second quantity is the population of higher order clusters, identified by the TCC (topological cluster classification) ~\cite{malins2013tcc}. Both observables are also influenced by the properties of the tracking routine in a way similar to that of a change in the state point. That makes the mapping somewhat subtle.

In general, we want to find the effective volume fraction of the experiments, the effective average hard sphere diameter of the particles and from that an estimate of the fraction of particles not being identified by the tracking algorithm. The mapping routine consists of 3 steps. The first step is a comparison of the total correlation function. The experimental data is binned onto the available simulation state points. The result is a first rough estimate of the effective volume fraction for the experiments.

The second step is a careful comparison of populations of specific higher-order clusters that have no similarities to crystal nuclei. In comparison to the pair correlation function, cluster populations are more sensitive to the state point and allow a more accurate second estimation of the effective volume fraction.

The last step deals with the fact that for experimental data we are only considering a subset of coordinates of the "true" system. We again compare the total correlation functions of experiments and computer simulations based on the results from the second step. From that comparison an effective hard sphere diameter $\sigma_\mathrm{eff}$ can be extracted that is used to calculate a second effective volume fraction from the number of particles tracked $N_{p}$.
\begin{equation}
\phi_\mathrm{exp}^\mathrm{counted} = N_{p} \frac{\pi \sigma^3}{6V}
\label{eqPhi}
\end{equation}
Here $V$ is the imaged volume of the experimental system. This second estimate is smaller compared to the true value ($\phi_\mathrm{exp}^\mathrm{counted} < \phi_\mathrm{exp}^\mathrm{true}$) due to particles lost during the tracking routine. To ensure that we actually compare the identical effective state points we delete the same fraction of particles from the simulation coordinates, so that $\phi_\mathrm{exp}^\mathrm{counted} = \phi_\mathrm{sim}^\mathrm{counted}$. The fraction of deleted particles varies slightly for different volume fractions around 1\%. A comparison of the distribution of those effective volume fractions is shown in the insets of Fig.~\ref{figAllBarriers}.

\section{Barriers}
\begin{figure}
\centering
\includegraphics[width=0.75\textwidth]{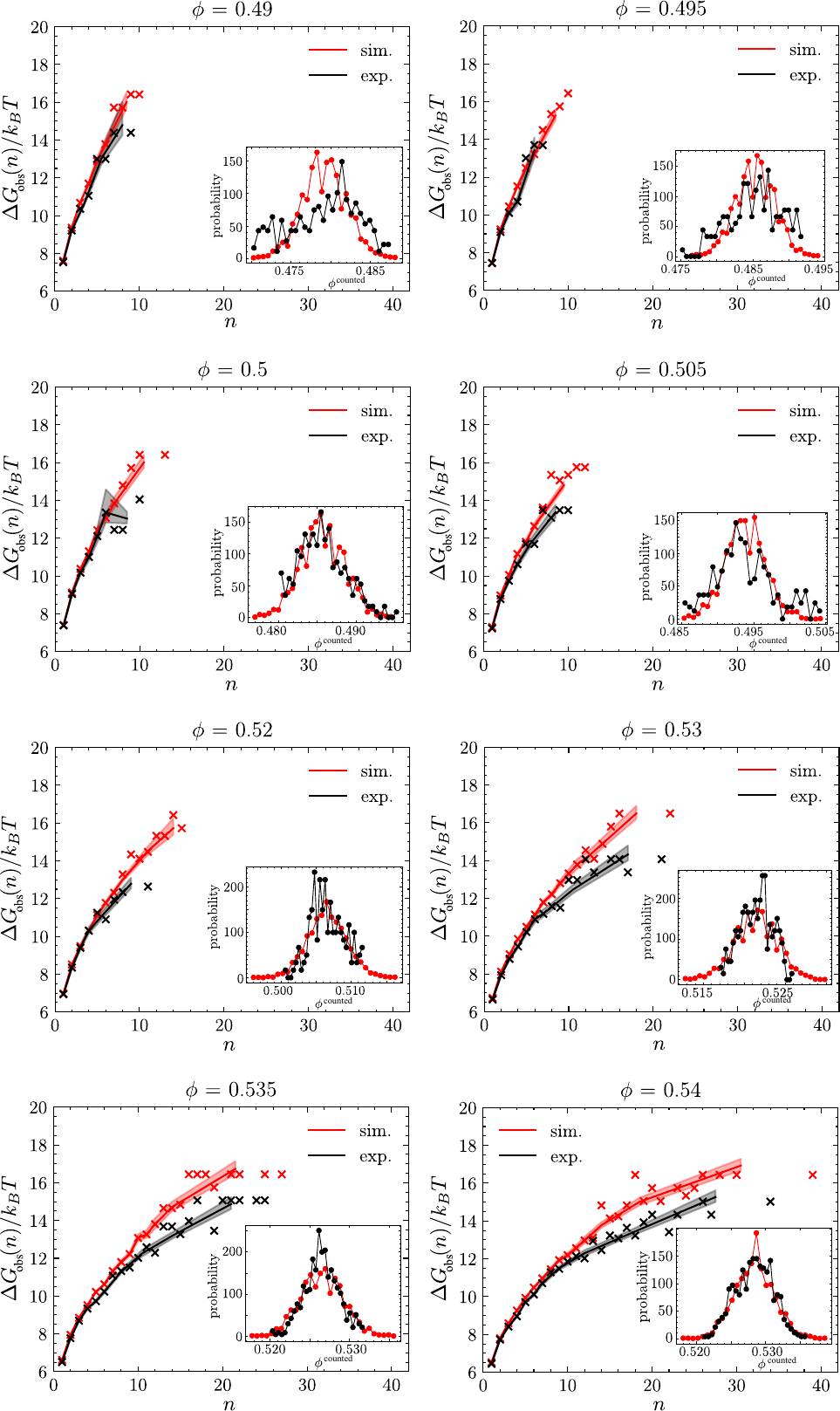}
\caption{\textbf{Effective Nucleation Barriers}. Raw output of the bond-orientational order parameter analysis (x-symbols) and the effective and binned nucleation barrier. Insets show a comparison of the effective volume fraction by counting particles between experiments and simulations.}
\label{figAllBarriers}
\end{figure}

The outcome of the bond-orientational order parameter analysis is a discrete value of the free energy barrier for a certain nucleus size. These discrete values were binned manually to ensure a reasonable representation of the raw data (Fig.~\ref{figAllBarriers}). The size of the bins varies according to the number of nuclei found for each size. The error bar is then calculated as the square root of the number of nuclei in each bin.

\begin{figure}
\centering
\includegraphics[width=.5\textwidth]{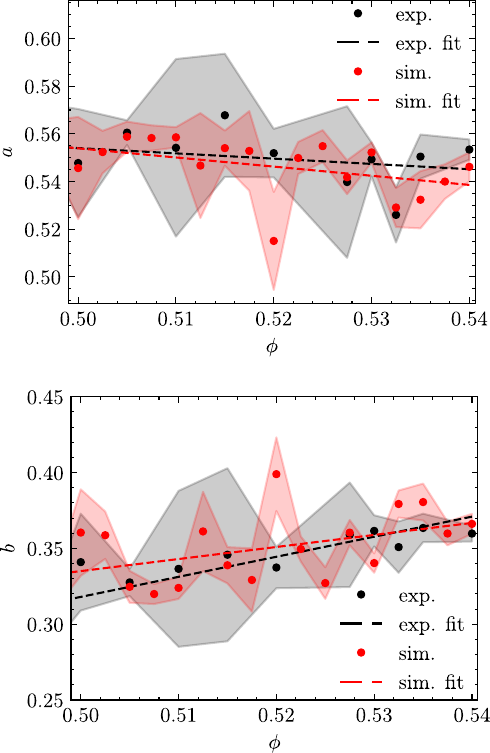}
\caption{\textbf{Fitting Parameters Radius of Gyration}. Parameters $a$ and $b$ of the power law fit $R_g(n)=a\cdot(n-1)^b$ for experiments and simulations as a function of the state point.}
\label{figPowerLaw}
\end{figure}

\end{document}